\documentclass[12pt,preprint]{aastex}
\usepackage{amsmath}

\received{}
\accepted{}
\journalid{}{}
\articleid{}{}
\paperid{}
\cpright{AAS}{2011}
\ccc{}

\shorttitle{Search for Higgs shifts in white dwarfs}
\shortauthors{Onofrio and Wegner}

\begin{document}

\title{Search for Higgs shifts in white dwarfs}

\author{Roberto Onofrio\altaffilmark{1,2} and Gary A. Wegner\altaffilmark{3}} 

\altaffiltext{1}{Dipartimento di Fisica e Astronomia ``Galileo Galilei,'' Universit\`a di Padova, 
Via Marzolo 8, 35131 Padova, Italy; onofrior@gmail.com}

\altaffiltext{2}{ITAMP, Harvard-Smithsonian Center for Astrophysics, 
Cambridge, MA 02138, USA}

\altaffiltext{3}{Department of Physics and Astronomy, 
Dartmouth College, 6127 Wilder Laboratory, Hanover, 
NH 03755, USA; gary.a.wegner@dartmouth.edu}

\begin{abstract}
We report on a search for differential shifts between electronic and vibronic transitions 
in carbon-rich white dwarfs BPM 27606 and Procyon B. 
The absence of differential shifts within the spectral resolution and
taking into account systematic effects such as space motion and 
pressure shifts allows us to set the first upper bound of astrophysical origin 
on the coupling between the Higgs field and the Kreschmann curvature invariant. 
Our analysis provides the basis for a more general methodology to derive 
bounds to the coupling of long-range scalar fields to curvature invariants 
in an astrophysical setting complementary to the ones available from
high-energy physics or table-top experiments. 
\end{abstract}


\section{Introduction}
The recent discovery at the Large Hadron Collider (LHC) of a resonance at 125 GeV compatible with 
the expectations for the Higgs particle \citep{Chatrchyan,Aad} represents a major step towards 
understanding the origin of the mass of fundamental particles. Eventually, this should 
also affect the other subfield in which mass has a pivotal role, {\it i.e.} gravitation. 
This is particularly relevant in models in which the Higgs field has nonminimal coupling to 
the general relativity sector, as invoked in various extensions of the standard model. 
Nonminimal coupling between the Higgs and spacetime curvature may be beneficial to have 
the Higgs responsible for inflation \citep{Bezrukov1,Bezrukov2}, and as a suppression 
mechanism for the contribution to dark energy expected from quantum fields \citep{Shapiro}. 
Upper bounds on the gravitational interaction of Higgs bosons from the LHC experiments 
have been recently discussed \citep{Atkins,Xianyu}.

Bounds on the crosstalk between the Higgs particle and gravity may also be obtained by 
considering strong-gravity astrophysical objects, as proposed in \citet{Onofrio} in the 
case of active galactic nuclei (AGN) and primordial black holes. The presence of a strong 
spacetime curvature deforms the vacuum expectaction value of the Higgs field and therefore 
the mass of fundamental particles such as the electron. Nucleons instead should be minimally 
affected by the strong curvature since most of their mass arises from the gluonic fields that, being 
massless, are not coupled to the Higgs field at tree level. Peculiar wavelength shifts are therefore predicted 
which should be present for electronic transitions and strongly suppressed for molecular transitions 
in which the main role is played by the nuclei themselves, such as in vibrational or rotational 
spectroscopy. Due to the vanishing of the Ricci scalar for spherically symmetric objects,  
attention was focused on the possibility of couplings to the only non-null curvature invariant, 
the Kreschmann invariant, defined as $K=R_{\mu\nu\rho\sigma}R^{\mu\nu\rho\sigma}$, where 
$R^{\mu\nu\rho\sigma}$ is the Riemann curvature tensor. This invariant plays an important 
role in quadratic theories of gravity \citep{Deser,Stelle,Hehl}, and more in general 
in modified $f(R)$ theories \citep{Sotiriou} and Einstein-Gauss-Bonnet models 
of gravity \citep{Lovelock}. 

While AGNs would provide a strong-gravity setting near their black holes, 
their complex structure and the presence of turbulence and high-energy interactions 
near the accretion region induce uncontrollable systematic effects which hinder the 
possibility for extracting bounds on a Higgs-Kreschmann coupling as this relies 
upon the simultaneous observation of atomic and molecular transitions. 
To our knowledge, no neutron stars appear to 
show both molecular and atomic lines in their spectra, while white dwarfs have both. 
Although their surface gravity is much weaker than around AGNs and neutron stars, 
many features can be controlled more precisely, thus providing a 'quieter' 
environment to search for the putative Higgs shift. 

White dwarfs have been known since the 19th century and in addition to their interest 
for astronomical and cosmological problems including understanding the late stages of 
stellar evolution, determining the galaxy's age, and the nature of Ia supernovae, they 
have had a prominent role in fundamental physics  since the early 20th century. 
\citet{Adams} made the first attempt to verify general relativity by measuring the 
gravitational redshift of Sirius B. 
\citet{Chandrasekhar} studied the consequences of Fermi-Dirac statistics for stars, 
introducing his celebrated limit. Bounds on the distance dependence of the Newtonian 
gravitational constant have been discussed comparing observations and models for the 
white dwarf Sirius B \citep{Hut} and those in the Hyades \citep{Wegner4}. 
More recently \citet{Berengut} proposed using white dwarfs to study the dependence 
of the fine structure constant on gravity. 
Here we show that white dwarfs can be used to obtain limits 
on the coupling of the Higgs field to a specific curvature invariant, by means of 
spectroscopic observations of a carbon-rich white dwarf, BPM 27606, using 
the Southern African Large Telescope (SALT). The analysis is complemented by considering 
data taken from the HST archive on a second white dwarf, Procyon B, in which CaII and MgII 
lines, in addition to the C${}_2$ bands, are also present.

\section{Higgs-shifts and Kreschmann invariant}

The search for coupling between the Higgs (or any scalar field permeating the whole Universe)
and spacetime curvature arises naturally within the framework of field theory in curved spacetime 
 \citep{Birrell}. The Lagrangian density for an interacting scalar field in a 
generic spacetime characterized by the metric tensor $g^{\mu \nu}$ is
written as \citep{Chernikov}:
\begin{equation}
{\cal L}= \sqrt{-g} \left[\frac{1}{2} g^{\mu \nu} 
\partial_\mu \phi \partial_\nu \phi - \frac{1}{2} (\mu^2 +\xi R) \phi^2 
-\frac{\lambda}{4}\phi^4\right],
\end{equation}
where $\mu$ and $\lambda$ are the mass parameter and the self-coupling quartic 
coefficient of the Higgs field, respectively. In Eq. 1 we have also introduced 
the determinant of the metric $g^{\mu\nu}$ as $g$, and $\xi$, the coupling 
constant between the Higgs field $\phi$ and the Ricci scalar $R$.
The coupling constant $\xi$ is a free parameter in any model so far imagined to describe 
scenarios of scalar fields coupled to gravity, and it is therefore important 
to extract this coefficient, or upper bounds, from phenomenological analyses.  
  
The Higgs field develops, under spontaneous symmetry breaking, a vacuum 
expectation value $v_0={(-\mu^2/\lambda)}^{1/2}$ in flat spacetime, and the 
masses of the fundamental fermions are proportional to $v_0$ via the Yukawa 
coefficients of the fermion-Higgs Lagrangian density term, $m_i=y_i v_0/\sqrt{2}$. 
The effective mass parameter of the Higgs field gets an extra-term due to the scalar 
curvature as $\mu^2 \mapsto \mu^2+\xi R$, and the vacuum expectation value 
of the Higgs field will become spacetime dependent through the curvature scalar as:
\begin{equation}
v=\sqrt{-\frac{\mu^2+\xi R}{\lambda}} \simeq v_0 \left(1+\frac{\xi R}{2 \mu^2}\right),
\end{equation}
where the approximation holds in a weak-curvature limit. This implies that the mass 
$m_i$ of fundamental fermions, such as the electron, will be simply changed proportionally 
to the Higgs vacuum expectation value 
\begin{equation}
\delta m_i=\frac{y_i}{\sqrt{2}} (v-v_0)\simeq 
\frac{y_i \xi R v_0}{2^{3/2}\mu^2} = \frac{\xi R}{2 \mu^2} m_i.  
\end{equation}
In other words, the presence of coupling of the Higgs field to
space-time curvature adds to its inertial mass a contribution, which 
acts as a 'mass renormalization' due to curved space-time (however, for an
opposite interpretation of this mass shift see \citet{Faraoni}). 
This mass shift is not present for protons and neutrons, due to 
the fact that their mass is primarily due to the gluonic fields which, being 
massless, are unaffected by the Higgs field. As discussed in more detail in 
\citet{Onofrio}, this implies that all molecular transitions only depending on the 
nuclei mass, such as vibrational and rotational spectra, should be unaffected by 
the Higgs-curvature coupling at leading order.

Unfortunately, the Ricci scalar outside spherically symmetric masses is zero, so 
we cannot use this coupling to infer possible mass shifts for the electrons. 
The only non-zero curvature scalar outside spherically symmetric masses is 
the Kreschmann invariant, and we will assume in the following considerations bounds 
to the Kreschmann coupling {\bf $\xi_K$} to the Higgs field in a Lagrangian of the form
\begin{equation}
{\cal L}= \sqrt{-g} \left[\frac{1}{2} g^{\mu \nu} 
\partial_\mu \phi \partial_\nu \phi - \frac{1}{2} (\mu^2 +\xi_K  \Lambda_\mathrm{Pl}^2 K)\phi^2 
-\frac{\lambda}{4}\phi^4\right],
\end{equation}
where $\Lambda_\mathrm{Pl}=(G\hbar/c^3)^{1/2}$ is the Planck length,
whose value is $\Lambda_\mathrm{Pl} \simeq 10^{-35}$ m in conventional
quantum gravity models, or larger values such as the one
corresponding to models with {\sl early} unifications of gravity to
the other fundamental interactions \citep{Arkani1,Arkani2}.  
In the latter case the Planck length occurs at the TeV scale via 
extra-dimensions, $\Lambda_\mathrm{Pl} \simeq 10^{-19}$ m, and  
in the following we will consider both these extreme situations.

Analogously to the case of the Ricci scalar, the mass parameter in the Higgs 
term then gets normalized as $\mu^2 \mapsto \mu^2(1+\xi_K \Lambda_\mathrm{Pl}^2 \lambda_{\mu}^2 K)$, 
where we have introduced the Compton wavelength corresponding to the Higgs mass, 
$\lambda_{\mu}=\hbar/(\mu c)$, equal to $\lambda_{\mu} \simeq 1.6 \times 10^{-18}$ m 
if we assume $\mu=125$ GeV.
Notice that, due to the subattometer scale values of $\lambda_{\mu}$ and 
$\Lambda_\mathrm{Pl}$, an extremely large value of $\xi_K$ is necessary for 
having mass shifts of order unity or lower to compensate for Kreschmann 
invariants due to macroscopic curvature of any spacetime of astrophysical interest. 
In fact, we get a relative mass shift, for instance in the case of the electron, equal to 
\begin{equation}
\frac{\delta m_e}{m_e} \simeq \frac{1}{2} \xi_K  \Lambda_\mathrm{Pl}^2 \lambda_{\mu}^2 K. 
\end{equation}
In the case of the Schwarzschild metric the Kretschmann invariant is 
$K=12 R_s^2/r^6$, with $R_s$ the Schwarzschild radius $R_s=2 GM/c^2$, 
and $r$ the distance from the center of the mass $M$. As a benchmark, for 
a solar mass white dwarf, $M= 1 M_\odot$ with a Earth radius 
$R=R_E \simeq 9.1 \times 10^{-3} R_\odot$, we get $R_s \simeq 3 \times 10^3$ m 
and $K \simeq 2 \times 10^{-33} {\mathrm{m}}^{-4}$. With $\Lambda_\mathrm{Pl} \simeq 10^{-35}$ m 
and the abovementioned value of $\lambda_{\mu}$ we obtain 
$\delta m_e/m_e \simeq 2.5 \times 10^{-139} \xi_K$ in MKSA units. 
For atomic transitions due to relocations of the electron in states with different 
principal quantum number, we expect that the mass shift affects the spectroscopy 
with a scaling of the transition wavelengths as 
$\delta \lambda/\lambda \simeq \delta m_e/m_e$, and therefore any evidence 
for a wavelength shift in the electronic transitions not accompanied by 
the same shift for transitions determined by the mass of the nuclei may be a distintive 
signature of Higgs-shifts. Therefore we need to detect emission or absorption wavelengths 
of both electronic and nuclear nature from a strong-gravity source, and make 
a comparison with either laboratory spectra or spectra gathered from weak-gravity 
astrophysical sources. 

The spectrum of the C${}_2$ molecule has been the subject of extensive 
experimental and theoretical studies in molecular spectroscopy, and has 
been found in various astrophysical contexts including white dwarfs.
In the visible region, the most prominent features of the C${}_2$ spectrum 
are the Swan bands, involving vibronic transitions between the electronic states 
d${}^3 \Pi_g$-a${}^3\Pi_u$ \citep{Huber,Brooke}. For these transitions, in the presence of 
a Higgs-shift the electronic energy levels, proportional to the electron mass, should be 
shifted, while the vibrational levels, proportional to the nucleon mass, should stay constant. 
We therefore expect that the separation between different terms of the same Swan band 
should stay constant, the only effect of the Higgs shift being an overall shift of all 
the wavelengths. In the following we therefore focus on these specific spectra as gathered 
from two white dwarfs.

\section{Observations and data analysis on BPM 27606}

In the Bruce Proper Motion Survey, \citet{Luyten} found 
BPM 27606\footnote{Other designations include: WD2154-512, GJ841B, L283-7, LDS765, LTT8768.}, 
which has a dMe common proper motion companion. \citet{Eggen} indicated that it is a white 
dwarf from {\it UBV} photometry. \citet{Wegner} discovered the strong C${}_2$
Swan bands which establish its spectral class as DQ in the current classification scheme,
and discussed its kinematical properties \citep{Wegner1}. More detailed descriptions 
of the spectrum were given by \citet{Wickr} and \citet{Wegner2}. Atmospheric 
analyses \citep{Koester,Wegner3} showed that BPM 27606 has a helium dominated atmosphere
(C:He $\sim 10^{-5}$) and effective temperature $T_{\mathrm{eff}} \sim 7,300$ K and more recent 
studies of DQ atmospheres ({\it e.g.} \citet{Dufour}) leave this essentially unchanged.

In recent years many new DQ white dwarfs have been found ({\it e.g.} \citet{Kleinman}) 
and several new properties about them have been discovered. These include rapid variability 
of some of the hot ones \citep{Williams}, rotation \citep{Lawrie}, and strong magnetic 
fields \citep{Williams}. An additional problem that remains unsolved is the physics 
of the blueshifts in the Swan bands of the DQs. These were already measured in the milder 
cases of BPM 27606 and L 879-14 \citep{Wegner2}, but for cooler DQs this becomes more 
extreme, such as for LHS1126 \citep{Bergeron}. \citet{Hall} and \citet{Kowalski} have 
reviewed possible mechanisms, but this subject is clearly hampered by the lack of information on 
the behaviour of the spectra of CI and C${}_2$ under white dwarf conditions.

BPM 27606 is a particularly good star for studying these effects because it is in the 
common proper motion system with CD-51${}^0$ 13128. This allows its true 
velocity to be known within fairly restrictive limits and the pair has a known 
distance from the trigonometric parallax. In addition, the shifts in the lines are relatively 
small ($\sim 4$ \AA) and its magnetic field is not large. \citet{Vornanen} have reported 
a magnetic field measurement of $1.3 \pm 0.5$ MG from circular polarization of the CH bands near 
$\lambda$4300. However the magnetic field may not be this high. 
From our new spectra this seems inconsistent with the lack of splitting of the 
$\lambda$4771 CI line and H$\alpha$, which indicates a magnetic field $B \le 2 \times 10^5$ G. 
Although variability is possible, the line is single on all of our spectra as it was in 
the 1978 spectra of \citet{Wegner2}.

The observations were made with the Robert Stobie Spectrograph (RSS) attached to the 
Southern African Large Telescope (SALT), which is described by \citet{Buckley}.
The RSS \citep{Burgh,Kobulnicky}) employs Volume 
Space Holographic transmission gratings (VPHGs) and three E2V44-82 2048 $\times$ 4096 CCDs 
with 15 $\mu$m pixels separated by gaps of 1.5 mm width, or about 
10-15 \AA ~ at the dispersions used here. 

All of the observations of BPM 27606 were obtained using a 1.0 arcsec $\times$ 8 arcmin slit 
rotated by 71${}^0$ so that both the white dwarf and its bright companion could be 
observed simultaneously. A 2 $\times$ 2 pixel binning with a gain of 1.0 e${}^-$/ADU gives 
a readout noise of 3.3 e${}^-$/pixel. The H${}_\alpha$ region ($\lambda\lambda$ 6085-6925) 
which also covers the $C_2(0,2)$ bandhead was observed 2013 May 2 with the pg2300 
grating and a pc04600 filter. Three 900 seconds exposures were obtained along with Ar 
comparison spectra. The FWHM measured from comparison lines is 1.3 \AA. 
An identical set of exposures were secured 2014 May 12 to improve the signal-to-noise ratio.
The $\lambda\lambda$ 4700-5380 wavelength region which covers the 5135 \AA 
~ bandhead was observed 2013 May 19 (two exposures) and 2013 May 21 (four exposures). 
All exposure times were 812 seconds with the pg3000 grating, no filter, and CuAr comparisons 
were used, giving a FWHM of comparison lines of 1.0 \AA. 
The 4737 \AA ~and 4382 \AA ~bandheads ($\lambda\lambda$ 4326-5004) were observed 2013 May 5 
and three 1042 second exposures using the pg3000 grating and no filter with a CuAr wavelength 
comparison. The resulting comparison line has a FWHM=1.1 \AA.
In 2014, two additional spectral regions were observed. Three 900 seconds exposures were 
secured 2014 May 13 of $\lambda\lambda$3540-4323 with the pg3000 grating, no filter and a 
ThAr comparison producing a FWHM resolution of 1.3 ~\AA~ from companion lines. 
Three 663 seconds exposures covering $\lambda\lambda$5050-6010 were made 2014 May 22 using the 
pg2300 grating with an Ar comparison giving a FWHM of 1.7 ~\AA~ for the comparison lines.

Starting from the bias subtracted and flattened images provided by the SALT pipeline 
\citep{Crawford}, the spectra 
were wavelength calibrated using the {\sc longslit} menu in IRAF\footnote{IRAF is distributed 
by the National Optical Astronomy Observatories which are operated by the Association of Universities 
for Research in Astronomy, Inc. under cooperative agreement with the National Science Foundation.}. 
Fifth order polynomials were used for the wavelength calibrations in two dimensions, background 
subtraction, and the spectra were extracted using {\sc apsum}. Special care was taken to use portions of 
the frames adjacent to the spectra, to minimize the effects of the curved lines in the RSS.
The final individual spectra were summed using {\sc imcombine} and the {\sc ccdsum} option 
to remove cosmic rays. Examples of the resulting spectra are shown in Figures 1-5.
The heliocentric velocity of the M2Ve companion, measured from our spectra using the 
H${}_{\alpha,\beta,\gamma}$ emission lines, was $-9.0 \pm 5.1$ km/s, which we adopt here.
This can be compared with $-9 \pm 1$ km/s \citep{Wegner1} and $-8.1 \pm 1.7$ km/s \citep{Karatas}.

The observed wavelengths of the major bandheads of C${}_2$ are shown in Table 1 where they are  
compared with the wavelengths given by \citet{Pearse}. These wavelengths refer to the minimum 
intensity at each bandhead from our spectra, and have been corrected to heliocentric values. 
We have analyzed the distance between two consecutive lines corresponding to the same $\Delta v=v'-v''$, 
for the observed wavelengths 
$\Delta \lambda_{\mathrm{obs}}(\Delta v)=\lambda_{\mathrm{obs}}(v')-\lambda_{\mathrm{obs}}(v'')$ 
and for the laboratory measured wavelengths
$\Delta \lambda_{\mathrm{lab}}(\Delta v)=\lambda_{\mathrm{lab}}(v')-\lambda_{\mathrm{lab}}(v'')$,  
evaluated both the overall average separation as 
$\langle \delta \lambda \rangle= N^{-1} \sum [\Delta \lambda_{\mathrm{obs}}(\Delta v)
-\Delta \lambda_{\mathrm{lab}}(\Delta v)]$ and its dispersion as 
$\delta \lambda^2=N^{-1} \sum [\Delta \lambda_{\mathrm{obs}}(\Delta v)-\Delta \lambda_{\mathrm{lab}}(\Delta v)]^2$,
obtaining the values of $\langle \delta \lambda \rangle= -0.71$ \AA ~ and $(\delta \lambda^2)^{1/2}=3.39$ \AA. 
The fact that $\langle \delta \lambda \rangle \ll (\delta \lambda^2)^{1/2}$ shows that, within the instrumental 
error, the spacing of the vibrational transitions is the same for the laboratory and the observed 
wavelengths. This provides a reliable anchor to study possible shifts in the bandheads due to the 
electron mass shift induced by the Higgs field. The stability of the vibrational transitions is 
further assured by their insensitivity to pressure-induced shifts since these are expected to be 
the same, in a linear approximation suitable for low pressures, for different vibronic bands \citep{Lin}.

\subsection{Assessment of temperature and surface gravity}

Due to the lack of recent detailed spectral scans of BPM 27606, its effective temperature 
is estimated from photometry. \citet{Koester} used 
intermediate band Str\"omgren uby data from \citet{Wegner1} and $\log g=8$ models 
which yield an effective temperature ${\mathrm{T_{eff}}} \sim 7,600$ K and $\log({\mathrm{He/C}}) 
\sim 4.6$. Using models of \citet{Galdikas1985} these colours give ${\mathrm{T_{eff}}} \sim 
7,200$ and  $\log({\mathrm{He/C}}) \sim 4.9$. 
\citet{Giammichele} found ${\mathrm{T_{eff}}}=(7,193 \pm 92)$ K,  $M = (0.60 \pm 0.07) M_{\odot}$, 
and $\log g=(8.03 \pm 0.04)$ in line, within few standard deviations, with our findings.
Broadband UBV photometry, 2MASS near infrared colours
\citep{Carollo} and an atmosphere model by \citet{Wegner2} bracket 
these values, so for the present estimates we adopt ${\mathrm{T_{eff}}}= 7,500$ K and 
$\log({\mathrm{He/C}}) \sim 4.75$.

To estimate the mass, we first determine the white dwarf radius using the relationship
$\log R/R_0=0.2 (V_0-V)-\log \pi+4.914$ with $V_0=-25.60$ from \citet{Koester}.
The visual magnitude $V=14.71$ is the mean of \citet{Eggen} and \citet{Bergeron}.
For the parallax we adopt $\pi=0.068 \pm 0.003$, the mean of three parallax measurements:  
from Hipparcos \citep{vanLeeuwen}, the Yale parallax catalogue \citep{vanAltena}, and 
the photometric parallax using the colour magnitude diagrams of Reid 
(http://www.stsci.edu \~ inr/cmd.html) for the dM companion.
This gives $R=0.0105 R_{\odot}$ which implies a mass $M \simeq 0.78 M_{\odot}$ according 
to the Hamada and Salpeter carbon or Chandrashekhar $\mu_e=2$ mass-radius relations. 
These values give a surface gravity of $\log g=8.3$ or $g=1.9 \times 10^8 {\mathrm{cm/s^2}}$. 
The corresponding gravitational redshift would be 
$V_{RS}=0.635 (M/M_{\odot})/(R/R_{\odot})=+47 ~{\mathrm{km/s}}$.

\subsection{Systematic effects}

The leading source of systematic effects in the C${}_2$ bands is pressure shifts in 
the dominant He background atmosphere, and in this section we estimate their order of magnitude. 
We interpolate in the DQ models of \citet{Galdikas,Galdikas1985} for the atmospheric parameters at 
Rosseland mean optical depth $\bar{\tau}=0.1$ for ${\mathrm{T_{eff}}}=7,500$ K and 
$\log (C/He)=-4.75$ for the order of magnitude of conditions in the line forming region.
This is $T_{0.1}=6,500$ K and $P_g=1.8 \times 10^9 {\mathrm{dynes~ cm^{-3}}}$. As the models use 
$g=10^8 {\mathrm{cm~ s^{-2}}}$, and for BPM 27606 $g=1.9 \times 10^8 {\mathrm{cm~ s^{-2}}}$, the 
number density is scaled to be $3.8 \times 10^{21}~ {\mathrm{cm^{-3}}}$. 

\citet{Hammond} measured pressure shifts for the five $C_2$ bandheads observed 
here in He under conditions close to those in BPM 27606 ($T=4,200$ K and 
$N=3 \times 10^{21} {\mathrm{cm^{-3}}}$). Although these measurements cannot explain 
the large blueshifts $\Delta \lambda \sim $ -200 ~ \AA ~ in the cooler peculiar DQ stars
\citep{Kowalski,Hall}, the shifts in BPM 27606 are much smaller than this and the 
laboratory measurements are of similar size, so it seems reasonable that for this star 
they can be used to estimate the pressure shifts. If the line broadening in the 
Swan bands resembles van der Waals broadening, the pressure shifts measured by 
\citet{Hammond}, $\Delta \lambda_{\mathrm{Ham}}$, would scale as 

\begin{equation}
\Delta \lambda_{\mathrm{press}}= 
\left(\frac{T_{0.1}}{4,200}\right)^{0.3} \left(\frac{N_{0.1}}{3 \times 10^{21}}\right) 
\Delta \lambda_{\mathrm{Ham}},
\end{equation}
with $T_{0.1}$ expressed in K and $N_{0.1}$ in $\mathrm{cm}^{-3}$.
For the estimated conditions at $\bar{\tau}=0.1$, 
$\Delta \lambda_{\mathrm{press}}=1.44 \Delta \lambda_{\mathrm{Ham}}$, which is 
used to correct the measured wavelengths of the $C_2$ bandheads given in Table 2 as 
$\Delta \lambda_{\mathrm{fin}}=\Delta \lambda_{\mathrm{meas}}-
\Delta \lambda_{\mathrm{press}}-\Delta \lambda_{\mathrm{GR}}$, where 
$\Delta \lambda_{\mathrm{GR}}$ is the gravitational redshift. 
The major source of systematic error in Equation 6 is the choice of $\bar{\tau}=0.1$ 
for the model atmosphere. This turns out to be relatively insensitive, as increasing the 
Rosseland optical depth to $\bar{\tau}=0.2$ would multiply $\Delta \lambda_{\mathrm{press}}$ 
in Tables 2 and 3 by a factor 1.2.

We also detect a weak feature at $\lambda=6563.33$ ~\AA. If this is the $H_{\alpha}$ line, 
$\lambda=6562.82$ ~\AA, no pressure shifts are expected, and this gives a gravitational 
redshift $V_{RS}=+52$ km/s. We do not detect the $H_{\beta}$ line in our spectra.
The CH(0,0) band at 4314.2 \AA ~ appears to be present as a weak feature at $\lambda4299$ and 
helps to confirm the presence of hydrogen.

A weak CI line can be seen in Figure 2 (see also Fig. 3 for details). 
The measured heliocentric wavelength neglecting pressure shifts is 4771.1 ~\AA. 
The laboratory wavelength of the 
strongest component of the $3s^{1}P^{0}-4p{}^3P$ multiplet is 4771.75 ~\AA ~ \citep{Moore}. 
There are currently no data on pressure shifts of CI lines.
The magnetic field on the surface of BPM 27606 is estimated looking at the 
resolution of the CI line. The classical Zeeman effect yields the  
relationship $\Delta \lambda = 4.7 \times 10^{-5} \lambda^2 B$ (in cgs units). 
The fact that only a single line with FWHM $\simeq \Delta \lambda=2.0$~ \AA ~ is 
visible (although with a low signal-to-noise ratio), yields an upper bound on 
the magnetic field of $B \leq 2 \times 10^{5}$ G, which makes negligible any 
correction to the estimates above. Using detailed calculations by 
\citet{Williams} of the magnetic splitting of this line also suggest such a low magnetic field.

The orbital motion is estimated considering the presence of the common proper motion 
companion to BPM 27606 with $V=10.49$, of spectral type M2Ve ($M=0.4 M_{\odot}$) at a 
projected separation of 28.45 arcsec, which at 15 pc yields a separation of $4.1 \times 10^2$ AU.
With a white dwarf mass of $0.8 M_{\odot}$ and taking this to be the semimajor 
axis of a circular orbit, the orbital period would be $P \sim 7.6 \times 
10^3$ years and the projected orbital velocity would be $\pm 0.5$
km/s, which is the order of magnitude of an additional source of uncertainty 
in the gravitational redshift determination. 

For BPM 27606 we conclude that an upper bound to the wavelength difference between the molecular 
C${}_2$ bands and the H${}_{\alpha}$ and CI atomic lines is 1.3 ~\AA. This is obtained by summing up 
the variance of the estimated pressure shifts (the square of the error in the second column of 
Table 2, corrected by the 1.44 Hammond scaling factor) and the square of the final difference between 
the processed and the laboratory wavelength (last column in Table 2), all other systematic errors 
being negligible with respect to these sources. 
If we disqualify the troubling CI line, it is 0.7 ~\AA. Although the corrections for pressure shifts 
in the C${}_2$ bands seems satisfactory, one must consider the uncertainties in their measurements 
which dominate the error budget in our estimate of the difference and is of the order of $\pm$ 2 ~\AA. 

\section{Observations and data analysis on Procyon B}

Procyon B is another white dwarf that has both atomic and molecular features 
in its spectrum and is in a well known binary. 
Its orbit and mass have been long studied ({\it e.g.} \citet{Schaeberle}; \citet{Spencer};
\citet{Strand}). \citet{Girard} obtained masses of $m_A=1.495 M_{\odot}$ and 
$m_B=0.602 M_{\odot}$ for the two components. Although Procyon B was long known 
to be a white dwarf, its spectrum could not be studied due to its proximity to its bright 
($V=$+0.34) primary (separation $\leq$ 5 arcsec). \citet{Provencal} secured spectra 
using the STIS instrument on the HST in 1998 February (Proposal 7398; PI H.L. Shipman) 
where the observations and data reductions are detailed. Here we adopt the atmospheric 
analysis in \citet{Provencal} which gives  ${\mathrm{T_{eff}}}=(7,740 \pm 150)$ K,
$R/R_{\odot}=0.0124 \pm 0.00032$ and  $\log C/He=-5.5$. These parameters show that 
Procyon B lies close to the carbon white dwarf mass-radius relation and predict a
surface gravity of  $g=1.1 \times 10^8 {\mathrm{cm~ s^{-2}}}$ and a gravitational 
redshift of $V_{RS}=+31 ~{\mathrm{km/s}}$. 

From the HST archives we used images 04g802010, 04g802020, 04g8020j0, and 
04g802090 which have suitable signal to noise ratio. We measured the 
wavelengths for $C_2$ bands, MgII and CaII lines. Figure 6 shows the $C_2$ bands near 
4,737 \AA ~and Figure 7 shows the CaII lines. The remainder of the spectrum is shown in 
\citet{Provencal}. The orbital velocity from \citet{Irwin} for the radial motion of 
Procyon B in 1998 is $-8.8 ~{\mathrm{km/s}}$. The models of \citet{Galdikas1985} using 
the ${\mathrm{T_{eff}}}$ and $g$ above indicate at $\bar{\tau}=0.1$, T${}_{0.1}$=6,750 K, 
and N${}_{0.1}=3.5 \times 10^{21} {\mathrm{cm^{-3}}}$.

The $C_2$ pressure shifts of \citet{Hammond} discussed in Section 3.2 are thus multiplied by 1.3.
The pressure shifts of the observed CaII and Mg lines produced by He are available and 
in both cases are redshifts. 
For CaII, theoretical values \citep{Monteiro} and laboratory measurements 
\citep{Hammond1989} agree relatively well. \citet{Monteiro} calculated 
pressure shifts for MgII and these were scaled to the above atmospheric parameters 
assuming van der Waals broadening. Table 3 summarizes the laboratory and measured 
wavelengths of features in the spectrum of Procyon B along with the corrections due to the 
orbital motion, $\Delta \lambda_{\mathrm{orbit}}$, the estimated pressure shifts 
$\Delta \lambda_{\mathrm{press}}$, and the gravitational redshift, $\Delta \lambda_{\mathrm{GR}}$. 
The resulting corrected wavelength of each line, $\lambda_{\mathrm{corr}}$, and the residual 
$\Delta \lambda_{\mathrm{fin}}=\lambda_{\mathrm{lab}}-\lambda_{\mathrm{corr}}$ are given. 
These are consistent with the results for BPM 27606.  By repeating the analysis as for the latter, and 
using the Hammond scaling factor of 1.3, the difference between atomic and 
molecular lines is formally 0.5 ~\AA, with the same uncertainties due to pressure shifts as before.

\section{Upper bounds to the Higgs-curvature coupling}

Based on the absence of relative shifts between the electronic and the vibrational 
transitions of the Swan bands, we are able to assess upper bounds on the Higgs-Kreschmann 
coupling. The expected wavelength shift is 

\begin{equation}
\frac{\delta \lambda}{\lambda} \simeq \frac{1}{2} \xi_K  \Lambda_\mathrm{Pl}^2 \lambda_{\mu}^2 K, 
\end{equation}
where the Kreschmann invariant is, for the estimated values of the mass and radius of BPM 27606,  
$K=12 R_s^2/R^6 \simeq 4.2 \times 10^{-34}~ \mathrm{m}^{-4}$. By assuming 
$\Lambda_\mathrm{Pl}=10^{-35}$ m, and the value of $\lambda_\mu$ quoted in Section 2, 
we obtain $\delta \lambda /\lambda \simeq 5.4 \times 10^{-140} \xi_K$, which may be inverted 
yielding an upper bound, for an average wavelength of $\lambda=5,000 ~{\mathrm{\AA}}$ 

\begin{equation}
\xi_K \leq 3.6 \times 10^{135} \left(\frac{\delta \lambda_{\mathrm{est}}}{1 {\mathrm{\AA}}} \right),
\end{equation} 
where $\delta \lambda_{\mathrm {est}}$ (in ~\AA) is the estimated wavelength resolution. 
If instead we use $\Lambda_\mathrm{Pl}=10^{-19}$ m as in models with extra dimension 
and quantum gravity at the Fermi scale \citep{Arkani1,Arkani2}, the bounds are a bit more 
constraining, as $\xi_K \leq 3.6 \times 10^{103}$ at $\delta \lambda_{\mathrm {est}}=1$ ~\AA. 

Both examples of upper bounds are expressed in meter-kilogram-second-ampere (MKSA) units 
of the Systeme International (SI), and it is worth converting them into natural units for 
the benefit of a comparison to the bounds already estimated from the observation of the 
Higgs particle at LHC \citep{Atkins,Xianyu}. The action term for the Higgs field including 
its coupling to spacetime, taking into account explicitly $\hbar$ and $c$, is expressed in SI units 
in terms of, for instance, the mass term as $S \propto \int d^3 x d(ct) \frac{m^2 c^2}{\hbar^2} \phi^2$.
Then the scalar field has dimensions $[\phi]=M^{1/2} T^{-1/2}$, and the Higgs-curvature term 
satisfying $S \propto \int d^3 x d(ct) \xi_{SI} \phi^2 R $ implies that $\xi$ is dimensionless. 
If the analysis is repeated for the action term expressed in natural units (NU), with 
$\hbar=1, c=1$, the dimensions of the scalar field change accordingly, and from the mass term 
it is simple to infer that $\phi_{NU}^2=(c/\hbar)^2 \phi^2$, implying the following relationship 
between the $\xi$ parameter evaluated in the two units systems, $\xi_{NU}=(\hbar/c)^2 \xi_{SI}=
1.23 \times 10^{-85} \xi_{SI}$. 
Analogous considerations may be repeated for the Higgs-Kreschmann coupling in which the curvature 
term has the same dimensions since $R$ is replaced by $\Lambda_\mathrm{Pl}^2 K$. 
This implies that in natural units the bounds on the Higgs-Kreschmann coupling 
become respectively, for the two extreme choices of $\Lambda_\mathrm{Pl}$, 
$\xi_K \leq 4.4 \times 10^{50}$ (for $\Lambda_\mathrm{Pl}=10^{-35}$ m) and 
$\xi_K \leq 4.4 \times 10^{18}$ (for $\Lambda_\mathrm{Pl}=10^{-19}$ m). 
This second bound on the Higgs-Kreschmann coupling is quantitatively comparable 
to the ones assessed on the Higgs-Ricci coupling through measurements at the LHC as
reported in \citet{Atkins} and \citet{Xianyu}, although in their analyses the 
usual $\Lambda_\mathrm{Pl}=10^{-35}$ m is assumed. With respect to bounds from 
table-top experiments based on tests of the superposition principle for gravitational 
interactions as discussed in \citet{Onofrio2012}, the bounds derived in this paper 
represent an improvement by ten orders of magnitude, as shown in Table 4. 

\section{Conclusions}

We have shown that observations in a somewhat controlled environment like the 
one provided by carbon-rich white dwarfs may be used to give upper bounds on the
coupling between the Higgs field and a specific invariant of the curvature of the 
spacetime, such as the Kreschmann invariant. The existence of nonminimal couplings 
between the Higgs and spacetime curvature is crucial to various proposals in which 
the Higgs also plays the role of the inflaton \citep{Bezrukov1,Bezrukov2}, and as 
a mechanism to suppress the dark energy contribution of quantum fields to the level 
compatible with the astrophysical observations based on SNIa \citep{Shapiro}. 

This methodology is complementary to the upper bounds recently discussed arising 
from the LHC experiments and their degree of agreement with the standard model of
elementary particle physics \citep{Atkins,Xianyu}, and can be adopted also to
search for couplings between generic scalar fields, not necessarily directly 
related to the Higgs vacuum, and space-time curvature, which may be competitive 
with bounds arising from the analysis of the cosmic background radiation as reported in 
\citet{Hwang,Komatsu1998,Komatsu1999}. Scalar fields, even if not directly interacting 
among themselves at level of their classical Lagrangian, will have a crosstalk once quantum radiative 
corrections are considered - the very origin of the hierarchy problem in Grand Unified Theories - 
so any scalar field will be then coupled  with the Higgs field and will indirectly affect   
the mass of elementary particles. Bounds based on this analysis could be even more stringent 
as many of the proposed candidates, for instance scalar fields invoked to accommodate the 
acceleration of the Universe \citep{Peebles,Ratra,Wetterich,Ostriker,Caldwell,Carroll,Bahcall,Wang} 
have a Compton wavelength greatly exceeding the one associated to the Higgs field, provided 
that compact astrophysical sources with non-zero Ricci scalar may be found.
 
The spectroscopic analysis presented here could be improved in a number of ways in the near future. 
Measurements and calculations of the pressure shifts of C I lines and the $C_2$ Swan 
bands under white dwarf conditions are needed. Observations of the ultraviolet 
carbon and other metal lines which are from lower energy levels would help disentangle the 
pressure shifts from the gravitational redshift measurement. 
A detailed scan of the white dwarf's spectral energy distribution combined 
with an updated atmosphere model would help understanding the details of 
the molecular and atomic carbon features. Additional objects of this type 
that are in binary pairs would help in the assessment of the local
gravity, further diversifying the sample to counterbalance peculiar
systematic effects.

\acknowledgments

We are grateful to Susanne Yelin for a critical reading of the manuscript.
Some of the observations reported in this paper were obtained with the 
Southern African Large Telescope (SALT) under proposals 2013-1-DC-001 and 
2014-1-DC-001. Additional data used observations made with the NASA/ESA Hubble 
Space Telescope, and were obtained from the Hubble Legacy Archive, which 
is a collaboration between the Space Telescope Science Institute (STScI/NASA), the Space Telescope 
European Coordinating Facility (ST-ECF/ESA) and the Canadian Astronomy 
Data Centre (CADC/NRC/CSA). This work was also partially funded by the National 
Science Foundation through a grant for the Institute for Theoretical 
Atomic, Molecular, and Optical Physics at Harvard University, and the 
Smithsonian Astrophysical Laboratory.

\newpage

\begin{table}
\begin{center}
\caption{\label{table1}
Observed heliocentric wavelengths of the Swan bands from BPM 27606 
for $v',v''$ transitions, and comparison to the laboratory 
wavelengths, all expressed in \AA ~units.}
\begin{tabular}{cccc}
\tableline
\tableline
$v',v''$ & $\lambda_{\mathrm{obs}}$ & $\lambda_{\mathrm{lab}}$ & $\lambda_{\mathrm{obs}}-\lambda_{\mathrm{lab}}$  \\
\tableline
0,2 & 6179.90 & 6191.2 & -11.3 \\
1,3 & 6114.74 & 6122.1 & -7.4 \\

\tableline
0,1 & 5630.35 & 5635.5 & -5.15 \\
1,2 & 5580.07 & 5585.5 & -5.4 \\
2,3 & 5536.00 & 5540.7 & -4.7 \\
3,4 & 5495.64 & 5501.9 & -6.3 \\
4,5 & 5467.73 & 5470.3 & -2.6 \\

\tableline
0,0 & 5159.97 & 5165.2 & -5.2 \\
1,1 & 5125.64 & 5129.3 & -3.7 \\
2,2 & 5094.45 & 5097.7 & -3.3 \\

\tableline

1,0 & 4733.05 & 4737.1  & -4.0 \\
2,1 & 4712.04 & 4715.2  & -3.2 \\
3,2 & 4694.74 & 4697.6  & -2.9 \\
4,3 & 4682.06 & 4684.8  & -2.7 \\
5,4 & 4675.53 & 4678.6  & -3.4 \\
6,5 & 4679.89 & 4680.2  & -0.3 \\

\tableline

2,0 & 4378.86 & 4382.5 &  -3.6 \\
3,1 & 4368.07 & 4371.4 &  -3.3 \\
4,2 & 4362.24 & 4365.2 &  -3.0 \\

\tableline
\end{tabular}
\end{center}
\end{table}

\begin{table}
\begin{center}
\caption{\label{table2}
Data processing for the extraction of the wavelength shift on BPM 27606, 
with all wavelengths expressed in \AA ~units.}
\begin{tabular}{ccccccc}
\tableline
\tableline
line 
& $\Delta \lambda_{\mathrm{Ham}}$  
& $\Delta \lambda_{\mathrm{press}}$  
& $\Delta \lambda_{\mathrm{meas}}$  
& $-\Delta \lambda_{\mathrm{comp}}$ 
& $\Delta \lambda_{\mathrm{GR}}$  
& $\Delta \lambda_{\mathrm{fin}}$ 
\\
(1) & (2) & (3) & (4) & (5) & (6) & (7) \\
\tableline
C${}_2$(0,2)          & -8.4 $\pm$ 0.3 & -12.1 & -11.3  & +0.18 & -1.0  & 0.0 \\
C${}_2$(0,1)          & -3.2 $\pm$ 0.5 & -4.6 & -5.15  & +0.17 & -0.9  & -1.3 \\
C${}_2$(0,0)          & -3.2 $\pm$ 0.5 & -4.6 & -5.2  & +0.16 & -0.8  & +0.0 \\
C${}_2$(1,0)          & -3.8 $\pm$ 1.0 & -5.5 & -4.0  & +0.14 & -0.7  & +0.9 \\
C${}_2$(2,0)          & -3.1 $\pm$ 1.0 & -4.5 & -3.6  & +0.13 & -0.7  & +0.3  \\
\tableline
CI($\lambda$4771) & -              & -    & -0.65 & +0.15 & -0.75 & -1.2  \\
H${}_{\alpha}$       & -              & -    &  +0.5 & +0.20 & -1.0  & -0.3 \\
\tableline

\end{tabular}
\end{center}
\tablecomments{Column (1) is the name of the line. Column (2) are Hammond's (1990) pressure 
shifts measurement for the C${}_2$ bands and their errors. Column (3) are the estimated pressure 
shifts in BPM 27606 using (2) and scaled as described in Section 3.2. Column (4) is 
the difference between the measured heliocentric wavelength from SALT spectra and laboratory wavelengths.
Column (5) is the correction for the system's motion based on the companion's radial velocity  $V_r=
-9.1$ km/s. Column (6) is the correction for a gravitational redshift of 47 km/s. Column (7) is 
the final difference between the processed white dwarf and laboratory wavelengths.}
\end{table}

\begin{table}
\begin{center}
\caption{\label{table3}
Data processing for the extraction of the wavelength 
shift on Procyon B, with all wavelengths expressed in \AA ~units.}
\begin{tabular}{cccccccc}
\tableline
\tableline
line  
& $\lambda_{\mathrm{lab}}$ 
& $\lambda_{\mathrm{obs}}$ 
& $\Delta \lambda_{\mathrm{orbit}}$  
& $\Delta \lambda_{\mathrm{press}}$ 
& $\Delta \lambda_{\mathrm{GR}}$ 
& $\lambda_{\mathrm{corr}}$ 
& $\Delta \lambda_{\mathrm{fin}}$ 
\\
(1) & (2) & (3) & (4) & (5) & (6) & (7) & (8) \\
\tableline
C${}_2$(0,0) & 5165.2  & 5161.8  & -0.15 & +4.2 & -0.53 & 5165.3 & +0.1 \\
C${}_2$(1,1) & 5129.3  & 5126.5  & -0.15 &      &       &        &      \\
C${}_2$(1,0) & 4737.1  & 4733.1  & -0.14 & +4.9 & -0.49 & 4737.4 & +0.3 \\
C${}_2$(2,1) & 4715.2  & 4714.5  & -0.14 &      & -0.49 &        &      \\
C${}_2$(3,2) & 4697.6  & 4696.1  & -0.14 &      & -0.48 &        &      \\
CaII H       & 3968.47 & 3969.84 & -0.12 & -1.4 & -0.41 & 3967.9 & -0.6 \\
CaII K       & 3933.66 & 3935.18 & -0.11 & -0.4 & -0.41 & 3934.3 & +0.7 \\
MgII         & 2802.70 & 2803.10 & -0.08 & -0.5 & -0.29 & 2802.2 & -0.5 \\
MgII         & 2795.53 & 2796.85 & -0.08 & -0.6 & -0.29 & 2795.9 & +0.4 \\
\tableline
\end{tabular}
\end{center}
\tablecomments{Column (1) is the name of the line. Column (2) is the 
laboratory wavelength. Column (3) is the measured heliocentric wavelength 
using the HST reduction. Column (4) is the wavelength correction due to the 
orbital motion of Procyon B. Column (5) gives pressure shift corrections 
described in Section 4. Column (6) is the correction for the gravitational 
redshift assuming that it is +31 km/s. Column (7) is the corrected wavelength 
of the observed line. Column (8) is the final difference between Procyon B and 
laboratory wavelengths,  $\Delta \lambda_{\mathrm{fin}}=
\lambda_{\mathrm{corr}}-\lambda_{\mathrm{lab}}$.}
\end{table}

\begin{table}
\begin{center}
\caption{\label{table4}
Summary of the upper bounds on Kretschmann-Higgs couplings $\xi_K$ (in natural units) 
from the analysis of the two white dwarfs (first two rows) and comparison to limits 
inferred from tabletop experiments on the validity of the superposition principle 
for gravitational interactions as discussed in \citet{Onofrio2012} (last row).  
Upper bounds to the Higgs-Kreschmann coupling constant $\xi_K$ have been 
evaluated for two different values of the Planck length as in the last two columns, 
the standard one and one assuming unification of gravity with the gauge interactions 
at the Fermi scale, respectively. The better wavelength resolution estimated for Procyon B 
is partly offset by the smaller value of its mass and the larger value of its radius with 
respect to BPM 27606, which affect significantly the bound due to the strong dependence
of the Kreschmann invariant on mass and radius.}

\begin{tabular}{cccccc}
\tableline
\tableline
System & $R/R_{\odot}$ & $M/M_{\odot}$ & $\delta \lambda_{\mathrm{est}}$ (\AA) & 
$\xi_k$($\Lambda_{\mathrm{Pl}}=10^{-35}$ m) & $\xi_k$($\Lambda_{\mathrm{Pl}}=10^{-19}$ m) \\
\tableline
BPM 27606             & 0.0105 & 0.78  & 1.3  & $5 \times 10^{50}$     & $5 \times 10^{18}$ \\
Procyon B             & 0.0124 & 0.602 & 0.5  & $9 \times 10^{50}$     & $9 \times 10^{18}$ \\
Table-top experiments &        &       &      & $2.5 \times 10^{60}$   & $2.5 \times 10^{28}$  \\
\tableline
\end{tabular}
\end{center}
\end{table}

\clearpage

\begin{figure}
\plotone{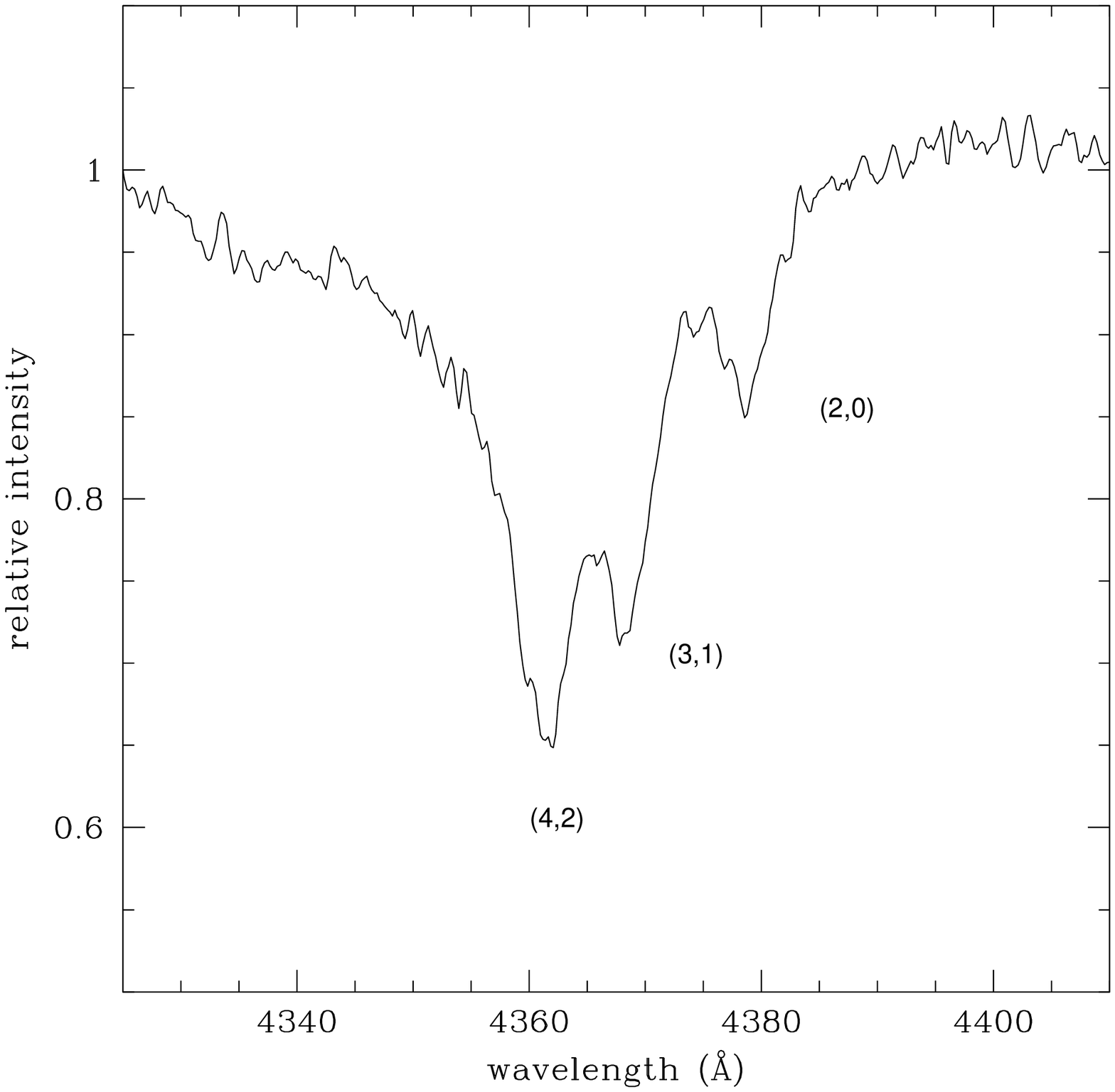}
\caption{Portion of SALT spectra of BPM 27606 showing the Swan band with $\Delta v=2$ of $C_2$ near 4370 \AA.}
\label{fig1}
\end{figure}

\begin{figure}
\plotone{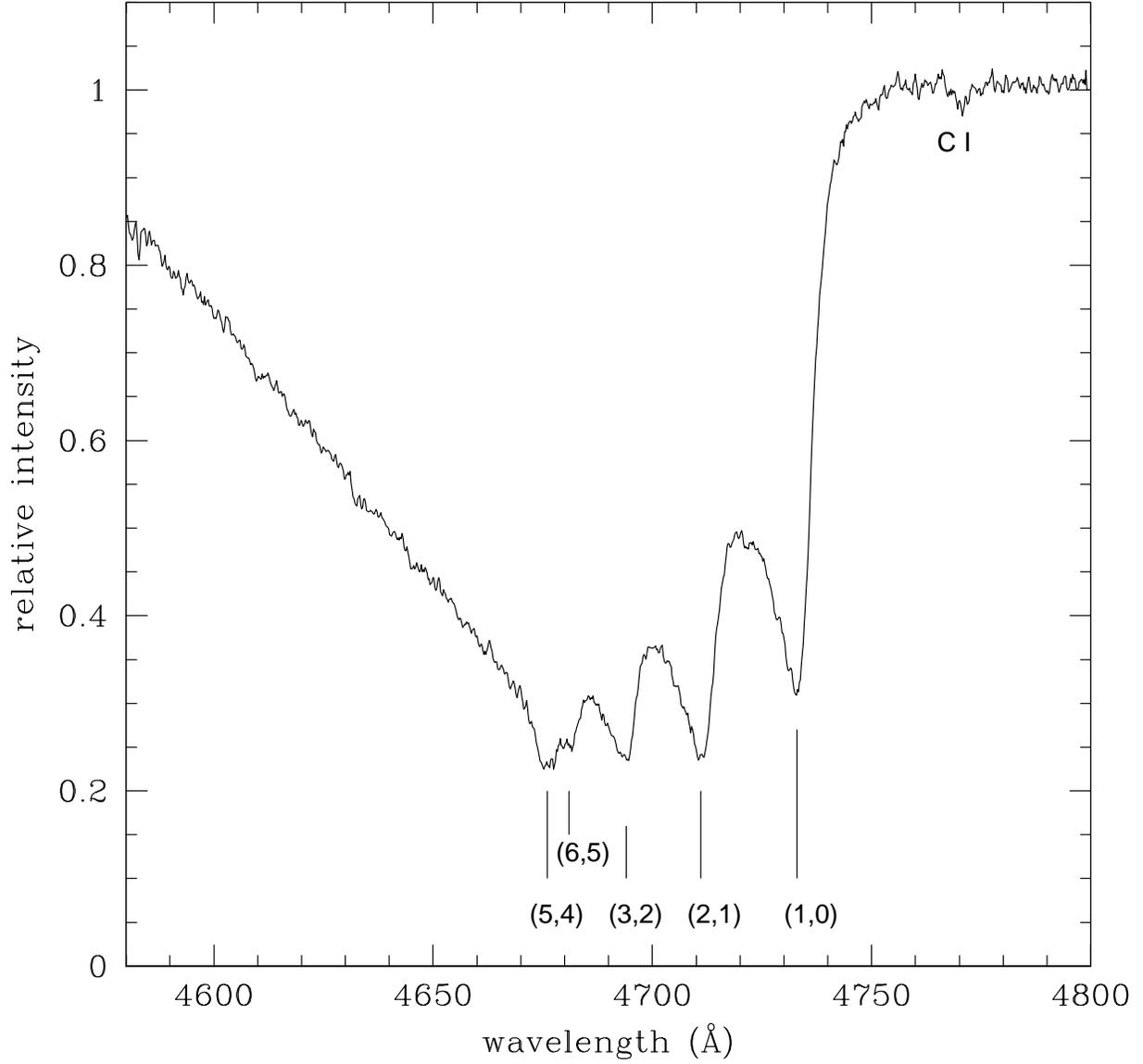}
\caption{Portion of SALT spectra of BPM 27606 showing the main Swan band with $\Delta v=1$ of $C_2$ and
the C I line near 4700 \AA.}
\label{fig2}
\end{figure}

\begin{figure}
\plotone{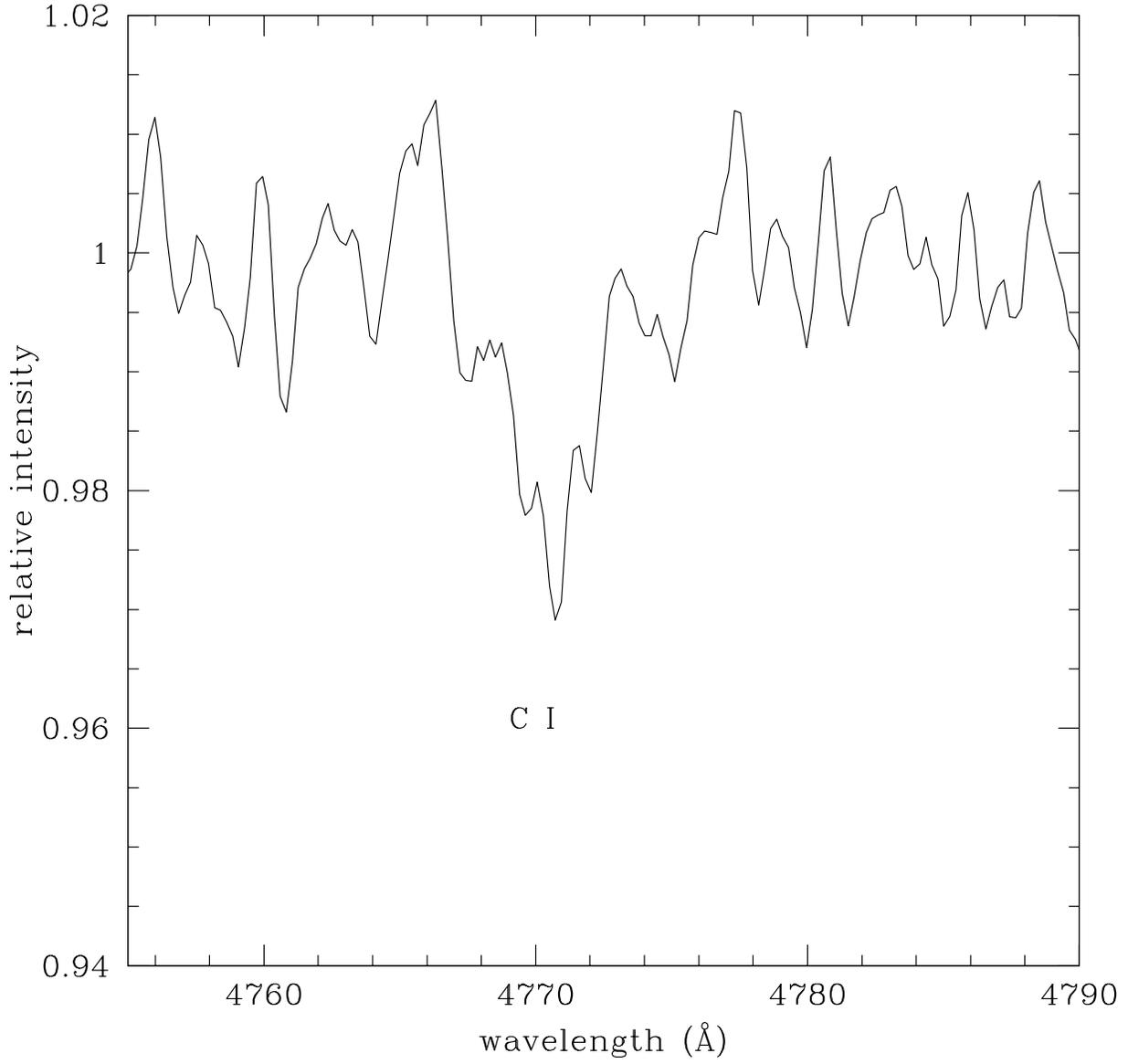}
\caption{Enlarged portion of the spectrum of BPM 27606 showing the neutral carbon line near 4770 \AA.}
\label{fig3}
\end{figure}

\begin{figure}
\plotone{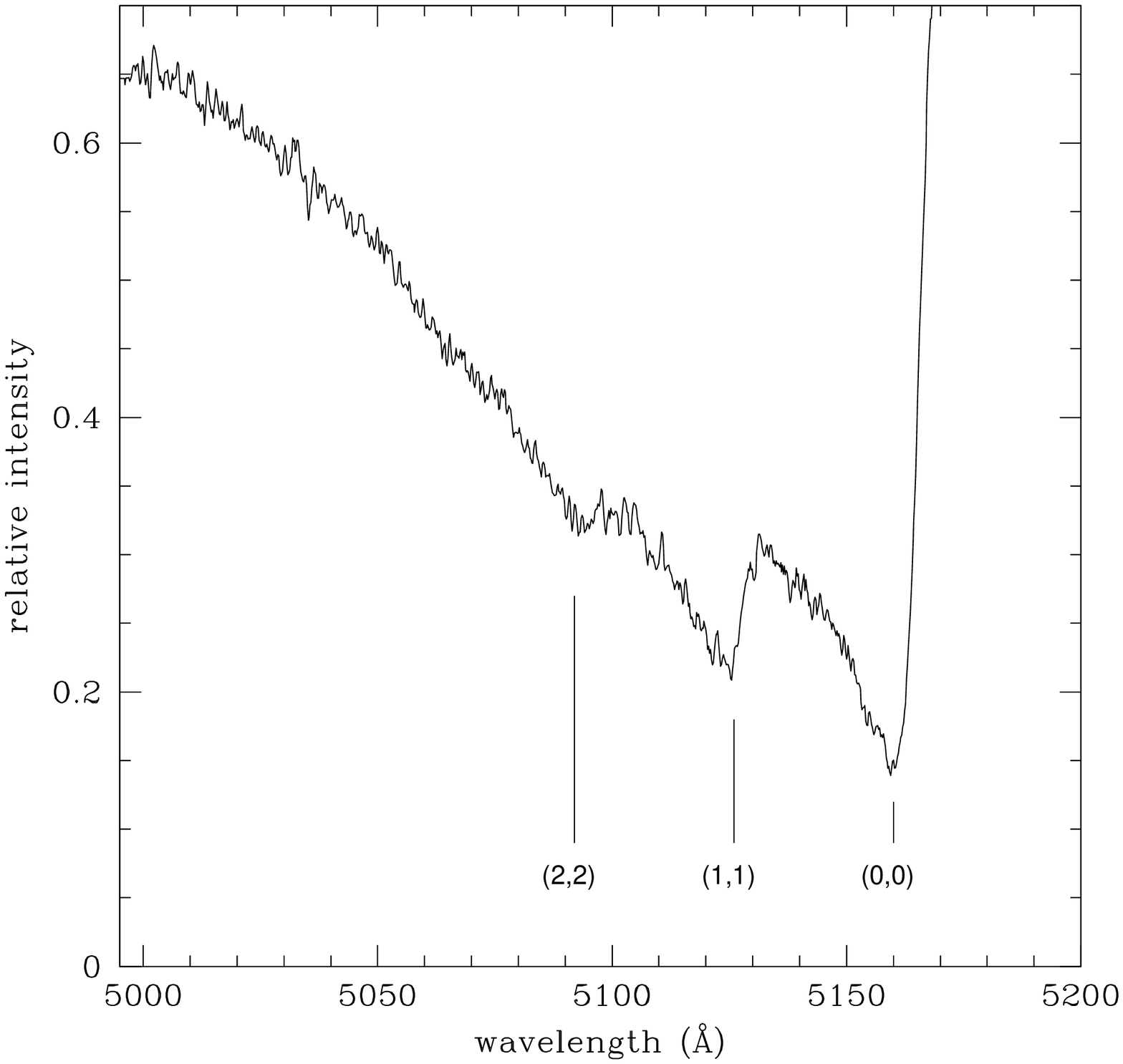}
\caption{Portion of SALT spectra of BPM 27606 showing the Swan band with $\Delta v=0$ of $C_2$ near 5100 \AA.}
\label{fig4}
\end{figure}

\begin{figure}
\plotone{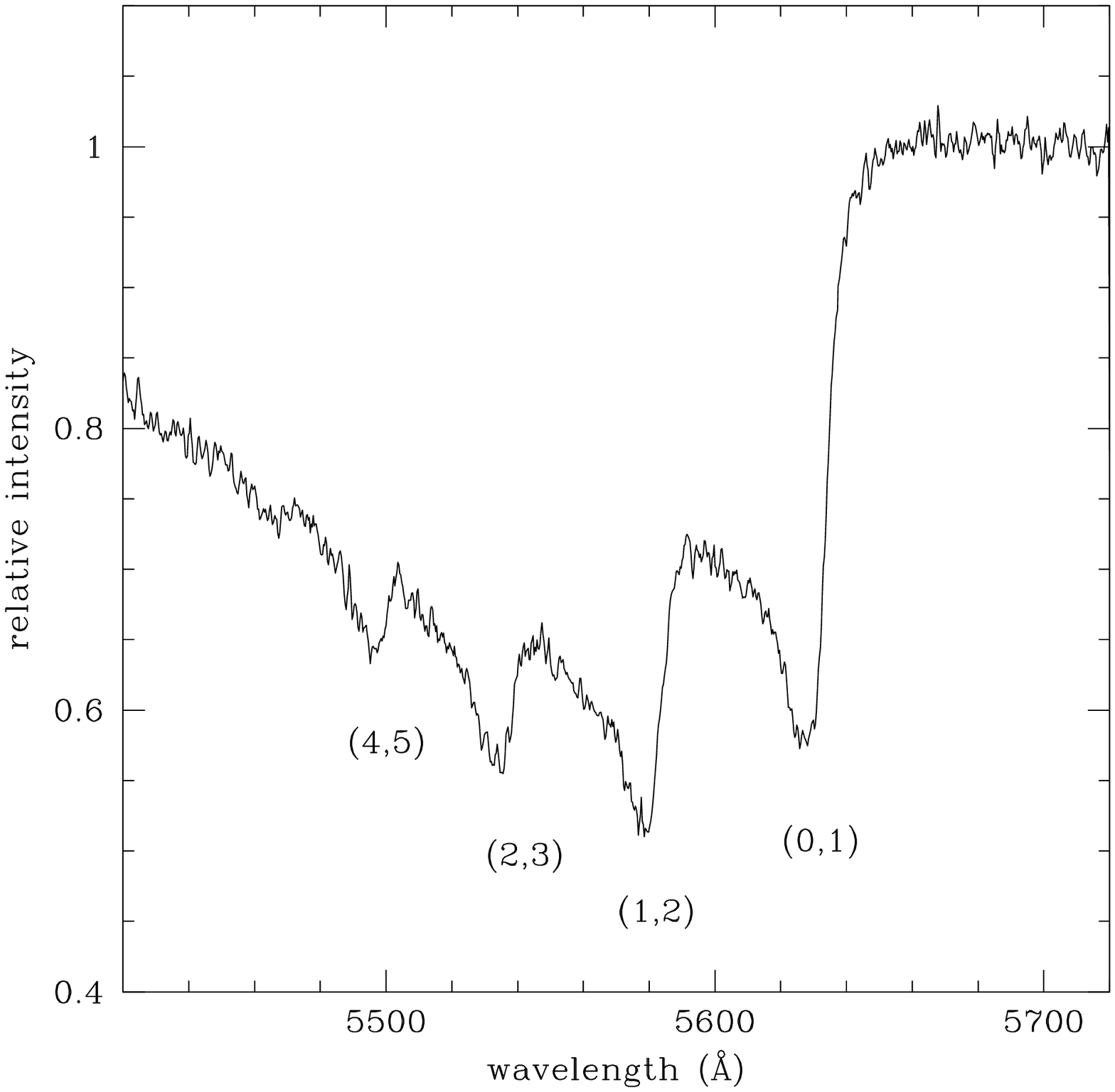}
\caption{Portion of SALT spectra of BPM 27606 showing the Swan band with $\Delta v=-1$ of $C_2$ near 5550 \AA.}
\label{fig5}
\end{figure}

\begin{figure}
\plotone{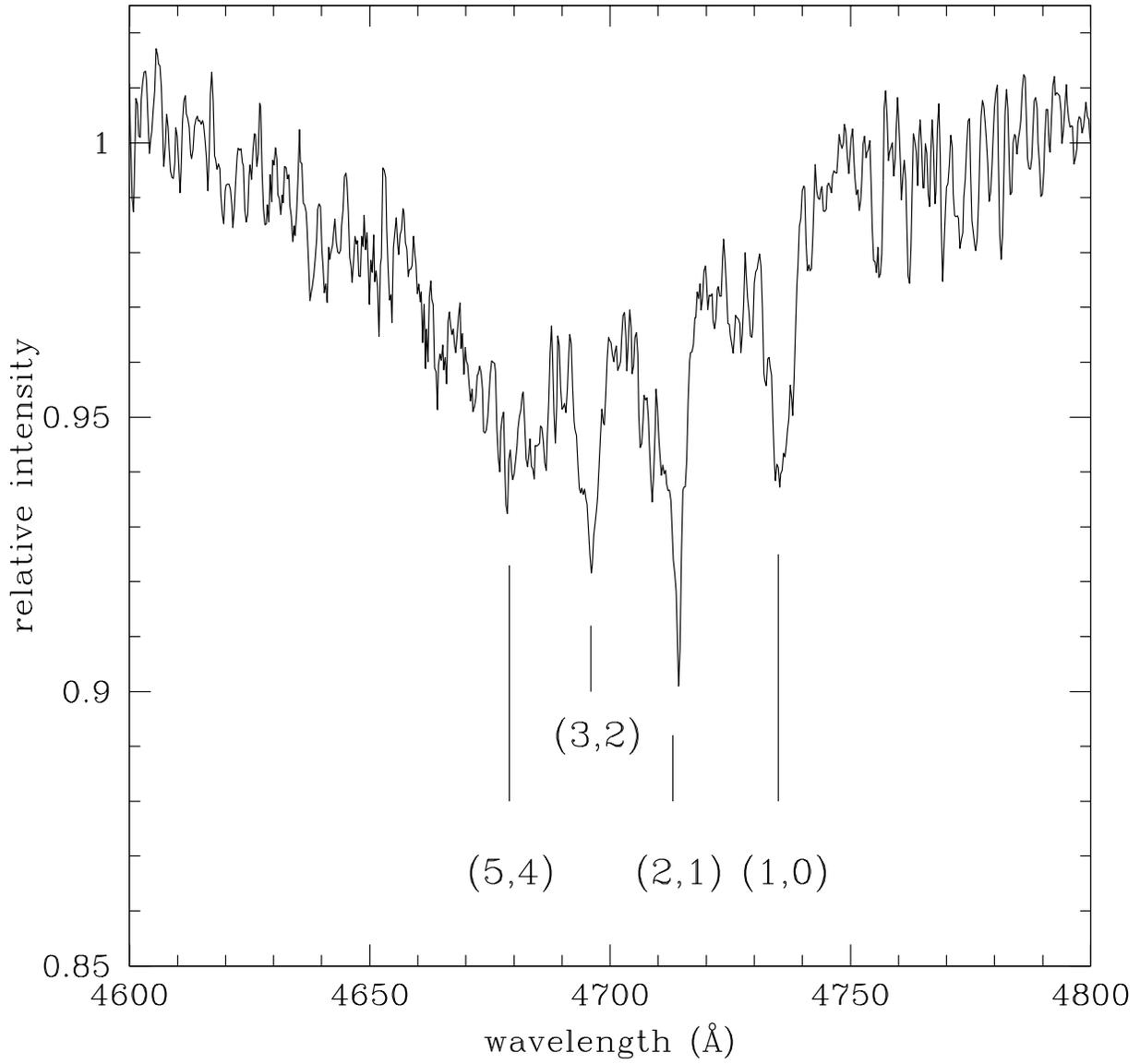}
\caption{Portion of spectra of Procyon B showing the C${}_2$ Swan band around 4700 \AA.}
\label{fig6}
\end{figure}

\begin{figure}
\plotone{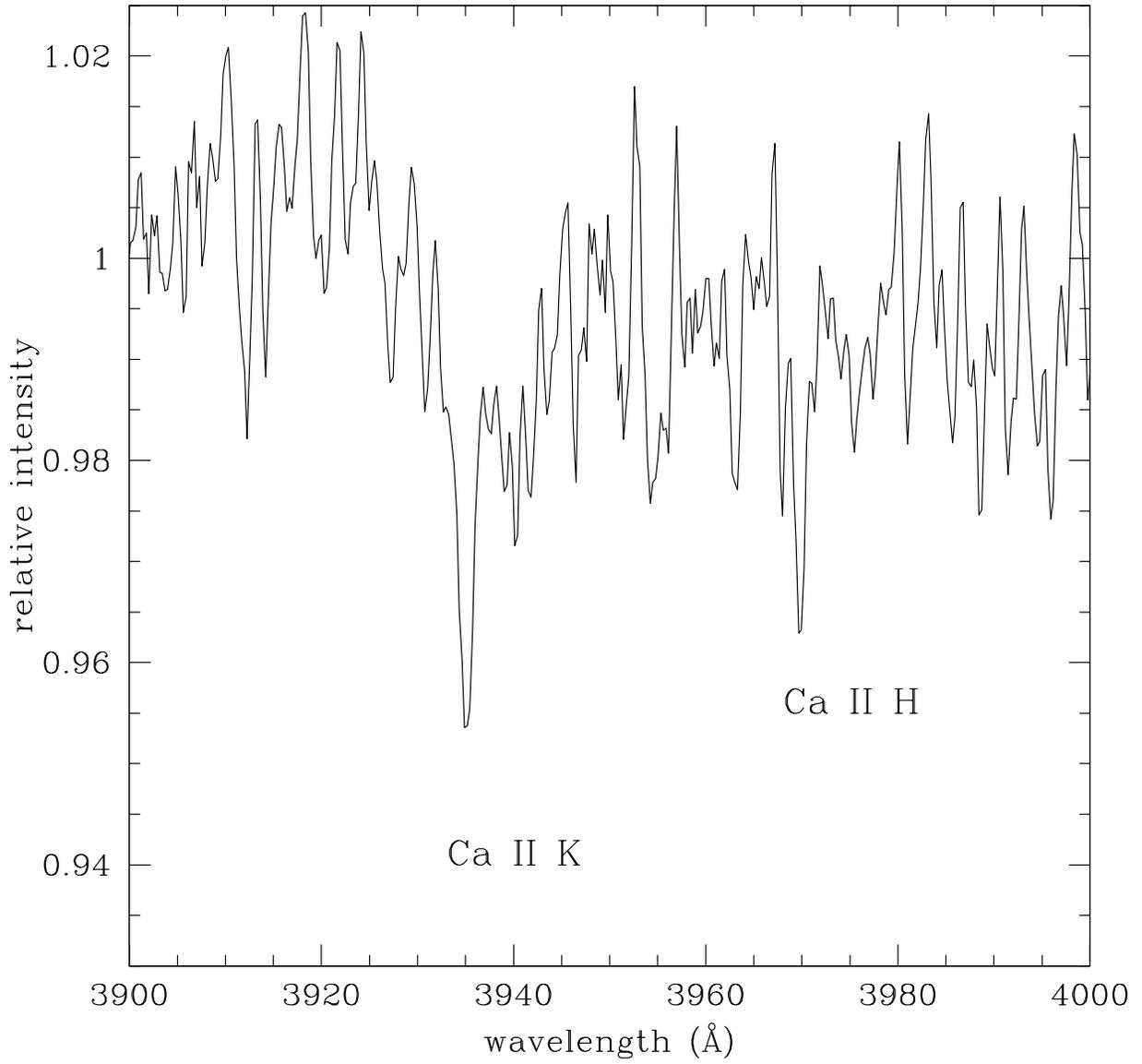}
\caption{Portion of spectra of Procyon B showing the Ca II H and K lines.}
\label{fig7}
\end{figure}

\end{document}